\documentclass{article}
\usepackage{spconf,amsmath,graphicx}
\usepackage{xcolor}
\usepackage{makecell,multirow,color,booktabs}
\usepackage{subcaption}
\usepackage{tabularx}
\usepackage{hyperref}

\title{Enhancing Spoofing Speech Detection Using Rhythm Information}
\name{Jingze Lu$^{1,2}$, Yuxiang Zhang$^{1,2}$, Wenchao Wang$^{1}$, Zengqiang Shang$^{1}$, 
Pengyuan Zhang$^{1,2,*}$\thanks{*Corresponding author}\thanks{ 
This work is supported in part by Youth Innovation 
Promotion Association CAS and by National Nature 
Science Foundation of China (12204509).}}
\address{$^1$Key Laboratory of Speech Acoustics and Content Understanding, Institute of Acoustics,\\Chinese Academy of Sciences, China\\
$^2$University of Chinese Academy of Sciences, China\\
\small \tt \{lujingze, zhangyuxiang, wangwenchao, shangzengqiang, zhangpengyuan\}@hccl.ioa.ac.cn}
\begin{document}
%
\maketitle
\begin{abstract}

Current spoofing speech detection systems need more convincing evidence. In this paper, the flaws of rhythm information inherent in the TTS-generated speech are analyzed to increase the reliability of detection systems. TTS models take text as input and utilize acoustic models to predict rhythm information, which introduces artifacts in the rhythm information. By filtering out vocal tract response, the remaining glottal flow with rhythm information retains detection ability for TTS-generated speech. Based on these analyses, a rhythm perturbation module is proposed to enhance the copy-synthesis data augmentation method. Fake utterances generated by the proposed method force the detecting model to pay attention to the artifacts in rhythm information and effectively improve the ability to detect TTS-generated speech of the anti-spoofing countermeasures.








\end{abstract}
\begin{keywords}
Anti-Spoofing Detection, Inverse Filtering, Generalization Ability, Logical Access
\end{keywords}
\section{Introduction}
\label{sec:intro}

Generating speech waveform with sufficient naturalness is a hot topic in the research field of speech signal processing. With the advent of various carefully designed text-to-speech (TTS)~\cite{shchemelinin2013examining} and voice conversion (VC)~\cite{kinnunen2012vulnerability} algorithms, synthesized speech is gradually becoming indistinguishable from bonafide one. While providing convenience for our lives, these utterances can also bring some potential security issues~\cite{wu2015spoofing}. Therefore, constructing robust and reliable speech spoofing detection algorithms is urgently needed.

Traditional features widely used in anti-spoofing countermeasures (CMs), such as Short-Term Fourier Transform (STFT)~\cite{zhang21da_interspeech} and Constant-Q cepstrum coefficients (CQCC)~\cite{todisco2017constant}, are considered adequate on some test sets. Features extracted by unsupervised pre-trained models can also be used for anti-spoofing tasks and achieve state-of-the-art performance~\cite{wang2021investigating,tak22_odyssey}. However, speech is the coupling of multi-dimensional features on a one-dimensional waveform. Although the features mentioned above might be capable of extracting most of the information from an utterance, detection models constructed with such features and Deep Neural Network (DNN) based back-ends do not provide convincing evidence. Such black-box CMs also trigger a crisis of trust in speech anti-spoofing systems in real-world scenarios. If classifiers cannot provide sufficiently 
interpretable evidence for judgment, why should we trust their results?

Therefore, a more in-depth analysis of synthesized speech is necessary. Currently, methods for generating fake speech could be roughly attributed to two categories: TTS and VC. In the paradigm of TTS algorithms, rhythm information is not provided directly. TTS models need to predict information such as rhythm, prosody, and fundamental frequency (F0) to generate waveform based on the provided textual content and speaker information. In contrast, the input of VC models consists of original waveform and speaker information. Generating waveform using VC algorithms is a speech-to-speech mapping, eliminating the need to generate rhythm information from scratch. Therefore, based on the differences in the processes of generating speech using TTS and VC algorithms, we hypothesize that rhythm information plays an essential role in detecting TTS-generated speech.

Based on this hypothesis, in this paper, we first investigate whether there are differences in rhythm information between bonafide speech and TTS-generated speech. Inspired by the process of human speech generation, we filter out the vocal tract response and achieve the glottal flow, which retains the rhythm information. Through experiments, we find that the anti-spoofing CMs can still distinguish TTS-generated speech training from rhythm information. Rhythm information plays an essential role in detecting TTS-generated speech. After demonstrating the hypothesis, we focus on utilizing rhythm information to enhance spoofing speech detection systems.

\begin{figure*}[tb]
  \centering
  \includegraphics[width=1.00\linewidth]{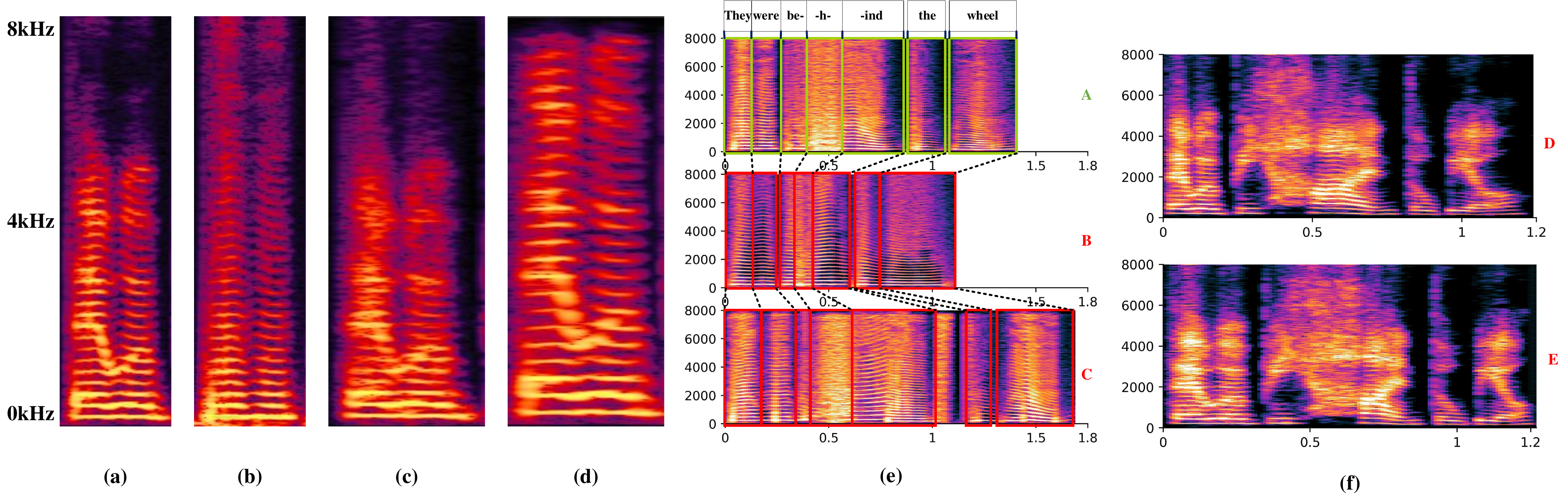}
  \caption{(a) Spectrum of a segment of a 
  \textcolor[HTML]{70AD47}{bonafide speech} \textcolor[HTML]{70AD47}{LA\_T\_4822766}
   from ASVspoof2019 dataset. (b) Spectrum of the glottal flow of 
   the same segment. (c) Spectrum of the same audio 
   segment extended to 1.5 times its original duration using 
   proposed rhythm perturbation method.
   (d) Spectrum of the same audio 
   segment extended to 1.5 times its original duration using speed perturbation method.
   (e) \textcolor[HTML]{70AD47}{A} is the complete spectrum of glottal flow of 
   \textcolor[HTML]{70AD47}{LA\_T\_4822766}, whose 
   textual content is \textbf{\textit{They were behind the wheel}}. \textcolor[HTML]{FF0000}{B}
    and \textcolor[HTML]{FF0000}{C} are 
   glottal flow of \textcolor[HTML]{FF0000}{TTS-generated} utterances, 
   with the same speaker and textual content as \textcolor[HTML]{70AD47}{A},
   named \textcolor[HTML]{FF0000}{LA\_T\_2224464} and 
   \textcolor[HTML]{FF0000}{LA\_T\_4184109}, respectively.
   (f) \textcolor[HTML]{FF0000}{D} is the spectrum of 
   copy-synthesis fake speech generated by WaveGlow vocoder~\cite{prenger2019waveglow}.
   \textcolor[HTML]{FF0000}{E} is generated by the same pipeline enhanced
   with RPM.}
  \label{fig:glottal}
\end{figure*}

Using pre-trained neural vocoders on bonafide utterances to generate fake utterances is a simple yet effective data augmentation paradigm called the copy-synthesis method~\cite{vocoder_copy}. However, such methods have limitations of their own. One of the limitations comes from the TTS acoustic model in modeling rhythm information. The pipeline of copy-synthesis does not require the prediction of rhythm information. The fake speech obtained by this method has the same rhythm as the natural speech. Anti-spoofing CMs trained with such a dataset could not pay attention to rhythm information, causing a performance decline for TTS-generated speech. Inspired by~\cite{pmlr-v119-qian20a}, we propose a rhythmic perturbation module to enhance the pipeline of copy-synthesis. It provides diverse rhythm information for the generated speech, guiding the classifier to focus on the rhythm artifacts of TTS-generated speech. In conclusion, our main contributions in this work include:
(1) We verify that rhythm information is vital in detecting TTS-generated fake utterances.
(2) We introduce a rhythm perturbation module to enhance the copy-synthesis data augmentation method paradigm and achieve performance improvements in detecting TTS-generated speech.

\vspace{-0.25cm}

\section{Method}
\label{sec:method}

This section first analyses whether there are differences in rhythm information between bonafide speech and TTS-generated speech. Subsequently, a method is proposed for enhancing the vocoder-based data augmentation method with a rhythm perturbation module.


\subsection{Rhythm Information of TTS-generated speech}
\label{ssec:subsubhead}

This work focuses on detecting TTS-generated speech using rhythm information. 
\textit{Rhythm} is a temporal representation that characterizes how fast the speaker utters each syllable. 
Due to the coupling of information, accurately representing rhythm information with a single hand-crafted feature is challenging. However, in the natural speech generation process, 
rhythm information is independent to a certain degree. The process of human speech generation is a coordinated workflow involving various organs. The source-filter model has approximated and simplified this process~\cite{fant1981source}. Through the compression of the lungs, airflow is generated. Upon passing through the glottis, airflow gives rise to glottal flow. The glottal flow undergoes modulation within the vocal tract, resulting in the production of initial speech. This speech subsequently radiates outward via the lips and oral cavity. In the frequency domain, the process could be expressed as $S(f) = G(f)V(f)L(f)$~\cite{perrotin2019spectral}, where $S$ and $G$ are the spectrums of speech and glottal flow. $V$ and $L$ are the vocal tract and lip radiation filters, respectively. 

In this generation process, the F0 and rhythm information are mainly provided by the glottal flow, while the vocal tract modulation provides speaker and semantic information. Therefore, rhythm features could be obtained by predicting the vocal tract information of speech and filtering it out. The response of the vocal tract could be estimated by employing linear predictive coding (LPC). The response of lips radiation could be approximated as a derivative filter with coefficient d close to 1 \cite{mcloughlin2016speech}, $L(z)=1-dz^{-1}$. The glottal flow of a speech utterance can be achieved by inverse filtering the vocal tract filters and lip radiation filters of each frame and subsequently overlapping them. We introduce Iterative Adaptive Inverse Filtering (IAIF)~\cite{alku91_eurospeech} method to implement this process and obtain the glottal flow.

Figure \ref{fig:glottal} (a) and (b) show the spectrum of original speech and its corresponding glottal flow. It could be observed that, compared to the original speech, the glottal flow lacks formant features. However, no significant changes are apparent on the temporal scale. This observation suggests that the glottal flow retains the rhythm information while eliminating most textual and semantic information. Figure \ref{fig:glottal} (e) shows the glottal flow of three utterances from the ASVspoof 2019 dataset. One of these utterances is bonafide, while the other two are generated by TTS algorithms. Despite sharing the same speaker and textual content, these utterances exhibit significant differences in rhythm information.

To further verify the significance of rhythm information in detecting TTS-generated speech, experiments are conducted on ASVspoof datasets. Original waveform and glottal flow are used to train an anti-spoofing system, respectively. Tabel \ref{tab_result_1} presents the results of the experiments. The results indicate a degradation in the overall performance of the anti-spoofing system when utilizing glottal flow compared to raw speech. However, this performance degradation is primarily attributed to the VC-generated utterances. The ability of classifiers to distinguish TTS-generated speech exhibits only a slight decline. Even after filtering out speaker and semantic information, the residual rhythm information still significantly distinguishes TTS-generated speech. Having demonstrated the importance of rhythm information, the subsequent focus of this work is to explore how anti-spoofing countermeasures could be enhanced using rhythm information.

\begin{table}[h]

  \caption{Performance of anti-spoofing CMs trained with 
  original waveform and glottal flow on ASVspoof 19LA 
  and 21LA evaluation dataset.
  The structure of anti-spoofing CMs is described in detail in section \ref{ssec:detail}.
  EER(\%) $\downarrow$ is used as metric.}
  \label{tab_result_1}
  \vspace{10pt}
  \centering
  \setlength\tabcolsep{2.4pt}
  \begin{tabular}{|c|c|c|c|c|c|c|c|}
    \Xhline{1.5pt}
    
    \multirow{2}{*}{\makecell[c]{\textbf{Training} \\ \textbf{Dataset}}} & \multirow{2}{*}{\textbf{Input}} 
    & \multicolumn{3}{|c|}{\textbf{19LA eval set}} 
    &\multicolumn{3}{|c|}{\textbf{21LA eval set}} \\

    \cline{3-8}
    &&\makecell[c]{\textbf{Total}}&\makecell[c]{\textbf{TTS}} & \makecell[c]{\textbf{VC}} & 
    \makecell[c]{\textbf{Total}}&\makecell[c]{\textbf{TTS}} & \makecell[c]{\textbf{VC}} \\

    \cline{1-8}
    \multirow{2}{*}{\makecell[c]{\textbf{{19LA}}\\\textbf{eval set}}}& 
    Waveform& 0.22& 0.15 & 0.57 & 1.30 & 1.00 & 2.22 \\
    \cline{2-8}
    & Glottal flow&2.23& 0.33 & 6.01 & 12.37 & 3.92 & 28.51 \\
    \Xhline{1.5pt} 
  \end{tabular}
\end{table}

\subsection{Rhythm Perturbation Module for Vocoder-based Fake Audio
Generation}
\label{ssec:subsubhead}
Using vocoders to generate synthetic speech is a simple yet effective method for constructing training sets for anti-spoofing countermeasures~\cite{vocoder_copy}. However, synthetic speech generated in this manner has a consistent rhythm with natural speech, making it challenging for classifiers to leverage rhythm information. To address this issue, inspired by~\cite{pmlr-v119-qian20a}, we introduce a rhythm perturbation module (RPM) to enhance the vocoder-based synthesized speech generation pipeline. The proposed module is a random sampling module.


\begin{figure}[tb]
  \centering
  \includegraphics[width=1.00\linewidth]{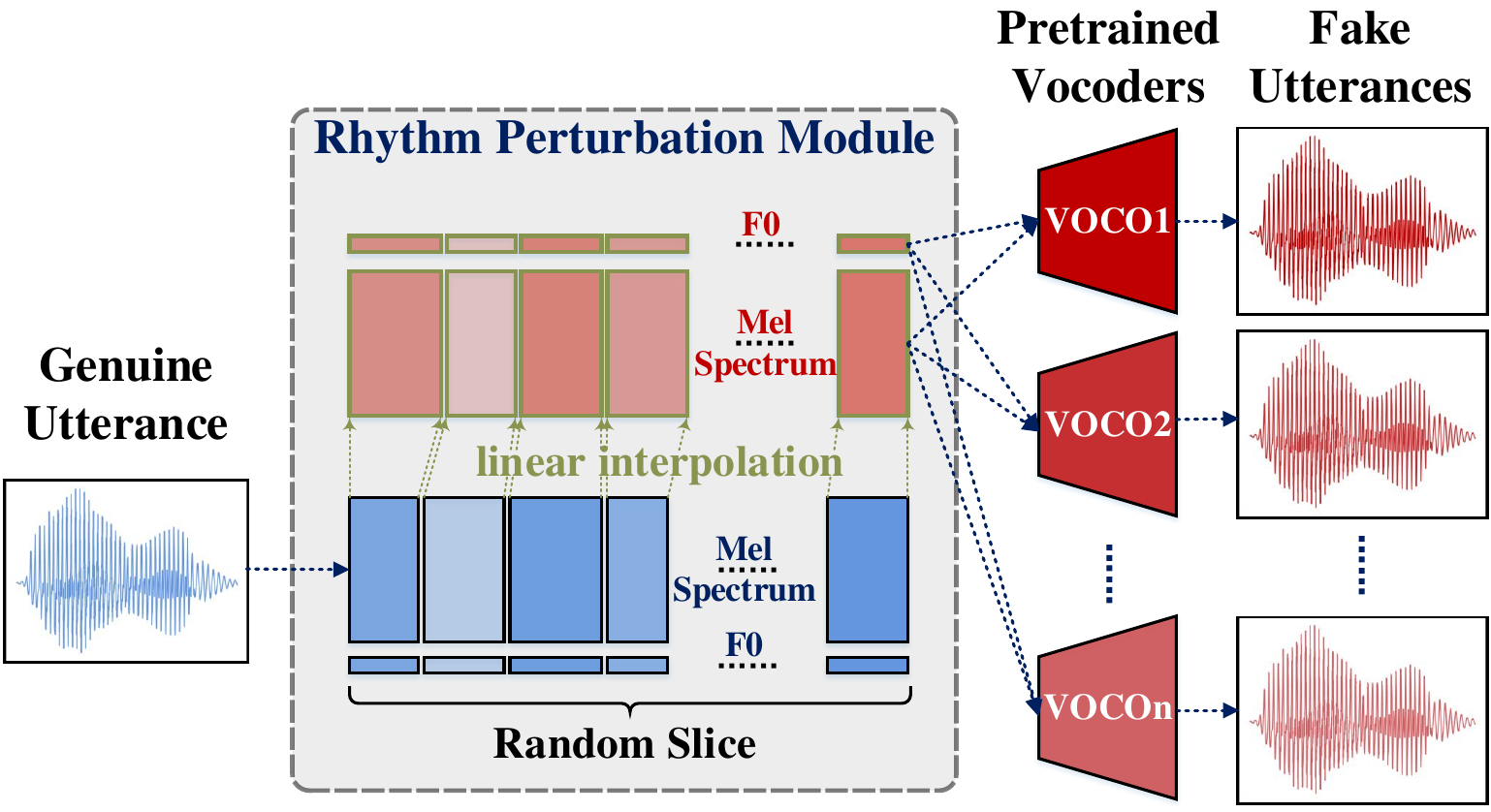}
  \caption{The pipeline of the proposed vocoder based 
  fake audio generating system with rhythm perturbation module.}
  \label{fig:pipeline}
\end{figure}

Figure \ref{fig:pipeline} shows the total pipeline of the proposed RPM-based vocoder-based fake audio generation system. For comparison, the vocoders provided by \cite{vocoder_copy} are utilized in this paper, all of which take mel-spectrum and F0 as input. In the vocoder-based copy-synthesis pipeline, the mel-spectrum and F0 of genuine speech are first extracted. Then, these features are divided into segments, whose length is randomly uniformly drawn from 19 to 32 frames. Each segment is resampled using linear interpolation with a resampling factor randomly drawn from 0.5 (compression by half) to 1.5 (stretch)~\cite{pmlr-v119-qian20a}. Finally, the segments are combined and fed into a pre-trained vocoder to generate synthetic speech. Figure \ref{fig:glottal} (f) shows the fake utterances generated with and without the rhythm perturbation module, and there exist significant differences between their rhythm. Compared to speed perturbation on segments, the proposed method does not change features corresponding to frequency. Figure \ref{fig:glottal} (c)(d) shows a comparison of the utterances obtained by the proposed method and speed perturbation. It could be observed that when the speech segment is stretched to 1.5 times its original length, speed perturbation causes both the F0 and harmonics of the speech to become 1.5 times their original values. In contrast, such issues do not exist with RPM.

\begin{table*}[tb]
  \caption{Performance of anti-spoofing CMs trained with fake utterances generated
  by the proposed RPM enhanced copy-synthesis method. 
  Voc.v3 and Voc.v4 are training sets generated with four different pre-trained vocoders~\cite{vocoder_copy}.
  The same pre-trained vocoders are utilized in the proposed RPM-based copy-synthesis method. 
  EER(\%) $\downarrow$ is used as metric.}
  \setlength\tabcolsep{2.8pt}
  \label{tab:dfresults}
  \small
  \centering
  \begin{tabularx}{\textwidth}{cc*{3}{X<{\centering}}|*{10}{X<{\centering}}|*{3}{X<{\centering}}}
    \Xhline{1.5pt}
    \multirow{2}{*}{\makecell[c]{\textbf{test}\\ \textbf{set}}}&
    \multirow{2}{*}{\makecell[c]{\textbf{training}\\ \textbf{set}}}&
    \multirow{2}{*}{\textbf{TTS}}&
    \multirow{2}{*}{\textbf{VC}}&
    \multicolumn{1}{c}{\multirow{2}{*}{\textbf{Total}}}&
    \multicolumn{10}{c}{\textbf{TTS attacks}}&
    \multicolumn{3}{c}{\textbf{VC attacks}}
    \\
    

    &&&&&\textbf{A07}&\multirow{1}{*}{\textbf{A08}}&
    \multirow{1}{*}{\textbf{A09}}&\multirow{1}{*}{\textbf{A10}}&
    \multirow{1}{*}{\textbf{A11}}&
    \multirow{1}{*}{\textbf{A12}}&
    \multirow{1}{*}{\textbf{A13}}&
    \multirow{1}{*}{\textbf{A14}}&
    \multirow{1}{*}{\textbf{A15}}&
    \textbf{A16}&
    \textbf{A17}&
    \multirow{1}{*}{\textbf{A18}}&
    \multirow{1}{*}{\textbf{A19}}
    \\
    \cline{1-18}
    \multirow{2}{*}{\textbf{19 LA}}&Voc.v4~\cite{vocoder_copy}&\textbf{1.84}&0.23&1.63&0.06&0.06&0.00&5.35&0.49&0.72&0.02&0.27&0.30&1.39&0.22&0.08&0.42\\
    &Voc.v4 RPM $_{0.5-1.5}$&2.22&0.90&2.00&0.30&0.02&0.02&3.82&0.47&1.36&0.08&0.22&0.42&1.10&0.20&0.18&1.49\\
    \cline{1-18}
    \multirow{6}{*}{\textbf{21 LA}}&Voc.v4~\cite{vocoder_copy}&14.68&7.07&13.72&10.76&8.06&5.29&29.09&15.52&14.51&7.95&14.11&10.25&16.88&3.90&4.77&9.35\\
    &Voc.v4 RPM $_{0.5-1.5}$&\textbf{10.43}&7.15&9.70&7.33&3.50&3.04&25.38&7.44&11.47&4.99&6.72&6.72&8.87&2.92&3.29&10.89\\
    &Voc.v4 RPM $_{0.7-1.3}$&11.06&7.28&10.06&5.94&2.64&2.39&26.34&6.47&13.37&4.65&6.71&7.24&11.54&2.59&2.61&10.25\\
    &Voc.v4 RPM $_{0.9-1.1}$&11.59&8.53&10.77&5.74&3.44&3.31&26.59&7.54&12.66&5.49&6.86&6.50&14.19&3.94&3.43&11.82\\
    \cline{2-18}
    &Voc.v3~\cite{vocoder_copy}&10.79&6.03&9.77&5.10&4.80&3.31&24.90&7.07&10.41&4.47&8.52&7.50&13.19&3.36&4.29&8.15\\
    &Voc.v3 RPM $_{0.5-1.5}$&\textbf{7.80}&5.47&7.18&2.75&1.46&0.93&26.92&3.76&6.99&1.61&4.33&5.40&4.05&1.47&1.61&8.87\\



    \Xhline{1.5pt}
\end{tabularx}
\end{table*}

\section{Experiments and Results}
\label{sec:results}

\subsection{Datasets and Metrics}
\label{ssec:dataset}

Experiments are conducted on ASVspoof datasets, which are influential datasets in spoofing speech detection. The ASVspoof 2019 logical access (19LA) dataset~\cite{wang2020asvspoof} is based on speech derived from the VCTK base corpus \cite{veaux2017cstr}. Fake utterances in the evaluation dataset are generated by 13 different TTS and VC systems, where $A07-A16$ are TTS attack models and $A17-A19$ are VC attack models. The ASVspoof 2021 LA (21LA) dataset~\cite{asv_21_towards} uses the same attack strategies as the 19LA dataset, while its utterances are transmitted over different various systems, including voice-over-IP (VoIP) and a public switched telephone network (PSTN). Only bonafide samples from the 19LA training set are used to construct our RPM training set. We follow the copy-synthesis method from~\cite{vocoder_copy}, and use the same pre-trained vocoders\footnote{\url{https://github.com/nii-yamagishilab/project-NN-Pytorch-scripts/tree/master/project/09-asvspoof-vocoded-trn}}. Equal error rate (EER) $\downarrow$ is used as the evaluation metric in this work.

\vspace{-0.15cm}

\subsection{Details of Systems Implementation}
\label{ssec:detail}
A pre-trained Wav2Vec 2.0 model~\cite{baevski2020wav2vec} 
is chosen as the front-end, and AASIST structure~\cite{jung2022aasist} 
is chosen as the back-end. 
The pre-trained Wav2Vec 2.0 model is optimized jointly 
with the AASIST back-end during the training process. 
All models are trained with Adam optimizer~\cite{kingma2014adam} 
with $\beta_1 = 0.9$, $\beta_2=0.98$, $\epsilon=10^{-8}$ 
and weigth decay $10^{-4}$. 
Angular margin based softmax loss (A-softmax)~\cite{Liu_2017_CVPR} 
is adopted as the loss function. 
The learning rate is fixed at $10^{-6}$. 
Training is conducted over 100 epochs and a batchsize of 32.






\vspace{-0.15cm}


\subsection{Result and Analysis}
\label{ssec:rpm}
Results are illustrated in Table \ref{tab:dfresults}. Baseline systems are trained using Voc.v4 and Voc.v3~\cite{vocoder_copy}. These two datasets consist of bonafide speech from 19LA and spoofing speech generated by four different pre-trained vocoders. The training datasets of Voc.v4 and Voc.v3 are different. Fake utterances generated by the RPM-enhanced copy-synthesis method are used to train a comparative system in which the same pre-trained vocoders are utilized. For the 19LA eval set, the EER of the RPM-based method is slightly decreased. The degradation is probably because the eval sets of 19LA are so clean that vocoder artifacts could be easily detected. Utterances from the 21LA dataset are transmitted over different codec algorithms, weakening vocoder artifacts in the spectrum. For the 21LA dataset, CMs trained with RPM-enhanced training data obtained significant performance improvement. Such performance enhancement is for the detection of TTS-generated speech. For all TTS-generated speech, CM trained on the Voc.v4 RPM $_{0.5-1.5}$ achieve an absolute improvement of $4.25\%$ in EER compared to the baseline. The proposed RPM method positively affects the detection of all ten types of TTS attacks $A07-A16$. RPM-enhanced fake data enables the detection system to focus more on the rhythm artifacts, thereby improving performance under complex channels. Meanwhile, for VC-generated speech, the performance remains essentially unchanged. This also validates our analysis of the significance of rhythm information for TTS-generated speech detection. The same experiments conducted on Voc.v3~\cite{vocoder_copy} and RPM-enhanced Voc.v3 obtain similar results.


In section \ref{ssec:subsubhead}, the resampling factor of RPM module is set to 0.5-1.5. In such a range, the rhythm of speech varies considerably. We change the resampling factor to 0.7-1.3 and 0.9-1.1 to explore the impact of different rhythm variation ranges on the CMs. The results indicate that RPM with different resampling factors can all enhance the ability to detect TTS-generated speech. A more extensive range of rhythm variations can yield better performance. It is worth noting that, no matter what resampling factor is chosen, the proposed RPM improves performance in detecting all ten TTS attacking algorithms.






\vspace{-0.25cm}

\section{Conclusion}
\label{sec:conclusion}

In this paper, we focus on enhancing the detection of spoofing speech 
using rhythm information. Rhythm characterizes the duration of each 
syllable in speech. 
TTS methods inherently require the prediction of rhythm 
based on text, which can result in rhythm artifacts. 
This paper first verifies a difference in rhythm information 
between TTS-generated speech and natural speech. 
The glottal flow retains rhythm information after 
filtering out the vocal tract response. 
Experiments demonstrate that a spoofing speech detection 
system trained with the glottal flow can still detect 
TTS-generated speech. 
Based on these analyses, a rhythm perturbation module is proposed 
to introduce rhythm artifacts to fake utterances generated by 
copy-synthesis methods. 
The proposed method significantly improves the detection of TTS-generated speech on the 21LA dataset. 
Our future work will focus on 
introducing rhythm information from TTS-generated speech into RPM.

\vfill\pagebreak



\bibliographystyle{IEEEbib}
\bibliography{strings,refs}

\end{document}